\documentclass[sigplan,screen, nonacm]{acmart}
\AtBeginDocument{%
  \providecommand\BibTeX{{%
    \normalfont B\kern-0.5em{\scshape i\kern-0.25em b}\kern-0.8em\TeX}}}

\acmConference[ArXiv]{ArXiv}{}{}
\acmPrice{}
\acmISBN{}

\newcommand\etal{\textit{et al.~}}

\newcommand{\metric}[1]{\lowercase{\textit{#1}}}

\newcommand{\targetnotation}[1]{\tilde{#1}}

\begin{document}

\title{Pose Metrics: A New Paradigm for Character Motion Edition}


\author{Léon Victor}
\orcid{0000-0003-1694-7189}
\affiliation{%
  \institution{Univ Lyon, INSA Lyon, LIRIS, UMR CNRS 5205}
  \country{France}
}

\author{Alexandre Meyer}
\orcid{0000-0002-0249-1048}
\email{alexandre.meyer@univ-lyon1.fr}
\affiliation{%
  \institution{Univ Lyon, Université Claude Bernard Lyon 1, LIRIS, UMR CNRS 5205}
  \country{France}
}

\author{Saïda Bouakaz}
\orcid{0000-0001-7091-7859}
\affiliation{%
  \institution{Univ Lyon, Université Claude Bernard Lyon 1, LIRIS, UMR CNRS 5205}
  \country{France}
  }

\renewcommand{\shortauthors}{Victor et al.}

\begin{abstract}
  In animation, style can be considered as a distinctive layer over the content of a motion, allowing a character to achieve the same gesture in various ways. Editing existing animation to modify the style while keeping the same content is an interesting task, which can facilitate the re-use of animation data and cut down on production time. Existing animation edition methods either work directly on the motion data, providing precise but tedious tools, or manipulate semantic style categories, taking control away from the user. As a middle ground, we propose a new character motion edition paradigm allowing higher-level manipulations without sacrificing controllability. We describe the concept of pose metrics, objective value functions which can be used to edit animation, leaving the style interpretation up to the user. We then propose an edition pipeline to edit animation data using pose metrics.
\end{abstract}





\maketitle

\section{Introduction}
Virtual characters are present everywhere in computer assisted media; populating virtual worlds, embodying players, telling stories. Producing qualitative motion for them is crucial to achieve engaging animation and raise interest in the viewer. Character animation software provide tools and interfaces which help artists do so. Many of such tools have been developed over the years, striving to find balance between precision, ease of use, and controllability. \cite{lasseterPrinciplesTraditionalAnimation1987, catmullProblemsComputerassistedAnimation1978} 

The human eye perceives a lot of information through motion: it is a complex behavior through which character convey intent, emotions or personality. A same gesture can be performed in an wide variety of ways, even by a same character. Editing an existing animation to change the way it is performed, but not the gesture itself, is then an interesting research topic. Many techniques have been proposed to tackle the problem of animation style edition, but most of them act from the highest level, and lack the fine-grained control users look for.

This paper presents a new framework for character motion edition using pose metrics as a middle-ground between raw bone manipulation and high-level "style" labels. 
We introduce these intermediate-level descriptors, then propose a neural-networks-based edition pipeline which allows animators to edit a character's motion by manipulating the metrics.

\section{Related work}

Virtual characters are traditionally represented by a mesh on which various textures and materials are applied. Its motion is driven by an underlying skeleton on which the mesh is fitted ("skinned"). The role of an animator is to design the character's motion by editing the skeleton's poses at specific time stamps, called key-frames. Instead of key-framing a new animation from the ground up this way, it is also possible to capture the motion of real-life actors through dedicated motion capture (MoCap) setups. MoCap data is a good source of realistic motion, but the resulting  animation still often needs to be cleaned up and edited to fit the animator's vision.   

Previous research on character animation investigates tools, algorithms and interfaces to help make authoring and editing motion a simpler process. 
Increased availability and quality of natural motion acquired by MoCap systems has led to raising interest in data-oriented methods. Meaningful information can be extracted from ground truth data and put to use to provide animators with more efficient tooling.


\subsection{Animation edition}
Whether originally key-framed by an an animator or recorded in a MoCap setting, animation data is set to change for any number of reasons, ranging from tweaking details to fit an artist's vision, to re-purposing it to be used on a different character. This paper builds upon a large body of work whose common goal is to provide animators with tools facilitating the edition process. The common animation edition pipeline can be split in two phases: posing and interpolation control. During the first, animators manipulate the skeleton to define key poses at specified key frames. In the second, they manipulate the interpolation process which is used to fill the gaps between key-frames.

\paragraph*{Posing.}
In most posing techniques the skeleton is conceptualized as a set of kinematic chains \cite{magnenat-thalmannJointDependentLocalDeformations1988}. With forward kinematics (FK), the modification of a joint's position or rotation trickles down, modifying joints down the chain. The opposite technique, inverse kinematics (IK) \cite{wangCCDInverseKinematics, aristidouFABRIKFastIterative2011}, find a suitable parameterization of the chain given a target end-effector, allowing the user to pose the character much like a puppet. These two methods are at the core of most modern animation software and lead to precise results, but can be tedious to use from an artist's perspective. Other techniques and interfaces have then been studied to alleviate this problem.

Some describe alternative frameworks in which the character's pose is derived from a line drawn by the user, much like a paintbrush stroke \cite{guayLineActionIntuitive2013, choiSketchiMoSketchbasedMotion2016}. This method has also been extended to virtual reality, tracking the user's hand in 3 dimensions and interpreting their motion as strokes \cite{garciaSpatialMotionDoodles2019}. On the data-oriented side, Wu \etal \cite{wuNaturalCharacterPosing2011} use learned clustering techniques to extract from a motion database a pose which most closely match the user's input. Victor \etal \cite{victorLearningbasedPoseEdition2021} train deep neural networks to solve the IK problem, learning skeleton constraints from existing data to lift some of the burden of respecting them from the user. Finally, some work has been dedicated to designing physical devices to facilitate the posing process \cite{yoshizakiActuatedPhysicalPuppet2011, jacobsonTangibleModularInput2014}.

\paragraph*{Interpolations.} Once key poses have been set, different methods exist to control how the software will interpolate between them to produce motion. Animators usually provide the interpolation factor used for each joint's position and rotation values over the elapsed time between key frames by manipulating animation curves through control points \cite{catmullClassLocalInterpolating1974}.
In the same veins, optimization-based methods have been proposed to improve the curve edition process \cite{koyamaOptiMoOptimizationGuidedMotion2018, cicconeTangentspaceOptimizationInteractive2019}. In the family of data-oriented solutions, Chai \etal \cite{chaiConstraintbasedMotionOptimization2007} propose to learn a low-dimensional space of motion, and solve the same kind of optimization problems within it. Recently, Harvey \etal \cite{harveyRobustMotionInbetweening2020} trained deep neural networks to learn to produce inbetweening motion from key frames.

Creating high-quality, appealing motion is a complex task. Animating a character according to fit one's vision, while depicting the desired action and conveying the right emotion requires an advanced knowledge of anatomy and physics rules. All the previously mentioned work allow animators to edit animation, but they all focus on controlling the character's body at bone-level. The process, while facilitated, is still a complex one. In order to provide animators and the greater public access to fast an expressive animation tools, higher-level concepts and tools are needed.  

\subsection{Motion style manipulation and transfer}

Stemming from the need of methods to reuse existing animation, another body of work has taken interest on editing animation by manipulating the motion style. The notion of \emph{style} is most commonly used as opposed to the notion of \emph{content}. Given an animation sequence, the terms are used as a way to separate \emph{what} the character is doing (the content) from \emph{how} it is doing it (the style). For example, several animations sequence could represent a character walking in various ways: fast, slow, making shorter or largers strides, swinging its arms... In this case they all share the same \emph{content}, walking, but the various \emph{styles} set them apart.

Being able to generate stylized animation, and to modify the style of an existing one, is then a desirable goal. Such control would allow the user to easily experiment with different styles, and it would be possible to generate a wide array of variations from an existing sequence.

Early solutions to the style edition problem use hand-made labels of style and content to categorize existing motion, and interpolate between categories to produce new motion. Rose \etal \cite{roseVerbsAdverbsMultidimensional1998} use such a method, coupled with an interface using a notion of verbs and adverbs to parameterize the generated motion. Brand \etal \cite{brandStyleMachines2000} describe a learned state machine, in which motion sequences are organized by shared style characteristics. More recently, Rodriguez \etal \cite{rodriguezParameterizedAnimatedActivities2019} stray away from the usual style paradigm and propose to edit animation trough high level trajectories parameterization.

Some other methods leverage the work produced in related fields, such as dance theory, to categorize motion style. Durupinar \etal \cite{durupinarPERFORMPerceptualApproach2016} build upon Laban Movement Analysis \cite{labanMasteryMovement1971} to attempt to formulate a link between personality and body motion. Aristidou \etal \cite{aristidouEmotionControlUnstructured2017} map the Laban features to a system of emotions coordinates, which can be interpolated in to edit and existing animation style.

Drawing inspiration from the success achieved by style transfer methods on images, Holden \etal \cite{holdenDeepLearningFramework2016} train a deep convolutional neural network to extract "style" features from of an animation and apply them to another. In the same vein, other types of neural networks have been used, such as variational autoencoders \cite{duStylisticLocomotionModeling2019}, adversarial learning \cite{wangAdversarialLearningModeling2020} or generative flow \cite{wenAutoregressiveStylizedMotion2021}.

On the opposite end of the spectrum from the posing methods discussed above, these style edition approaches share the common drawback of being \emph{too} high-level. Because the notion of animation style is still loosely defined (and this shortcoming is discussed further in the next section), tools using styles as control parameters could be met with incomprehension on the user's end when reaching broader audiences.

\section{Problem formulation}

The previously defined usage of \emph{style} as a separation from an animation's \emph{content} is commonly accepted in the computer animation literature. The limits of what the term itself entails are however not clearly defined.

Some associate an animation's style to emotions experienced by the character, such as "joy", "anger" or "sadness" \cite{fouratiEmilyaEmotionalBody2014}. Some link it to existing motion analysis systems, such as Laban's \cite{labanMasteryMovement1971, durupinarPERFORMPerceptualApproach2016}, or to psychological concepts such as the arousal/valence scale \cite{ekmanHeadBodyCues1967, randhavaneIdentifyingEmotionsWalking2019}. Some others use arbitrary categories, such as "gorilla", "childish", "zombie", or "neutral" \cite{xiaRealtimeStyleTransfer2015}. 

\begin{figure}[h!]
    \centering
    \includegraphics[width=0.8\columnwidth]{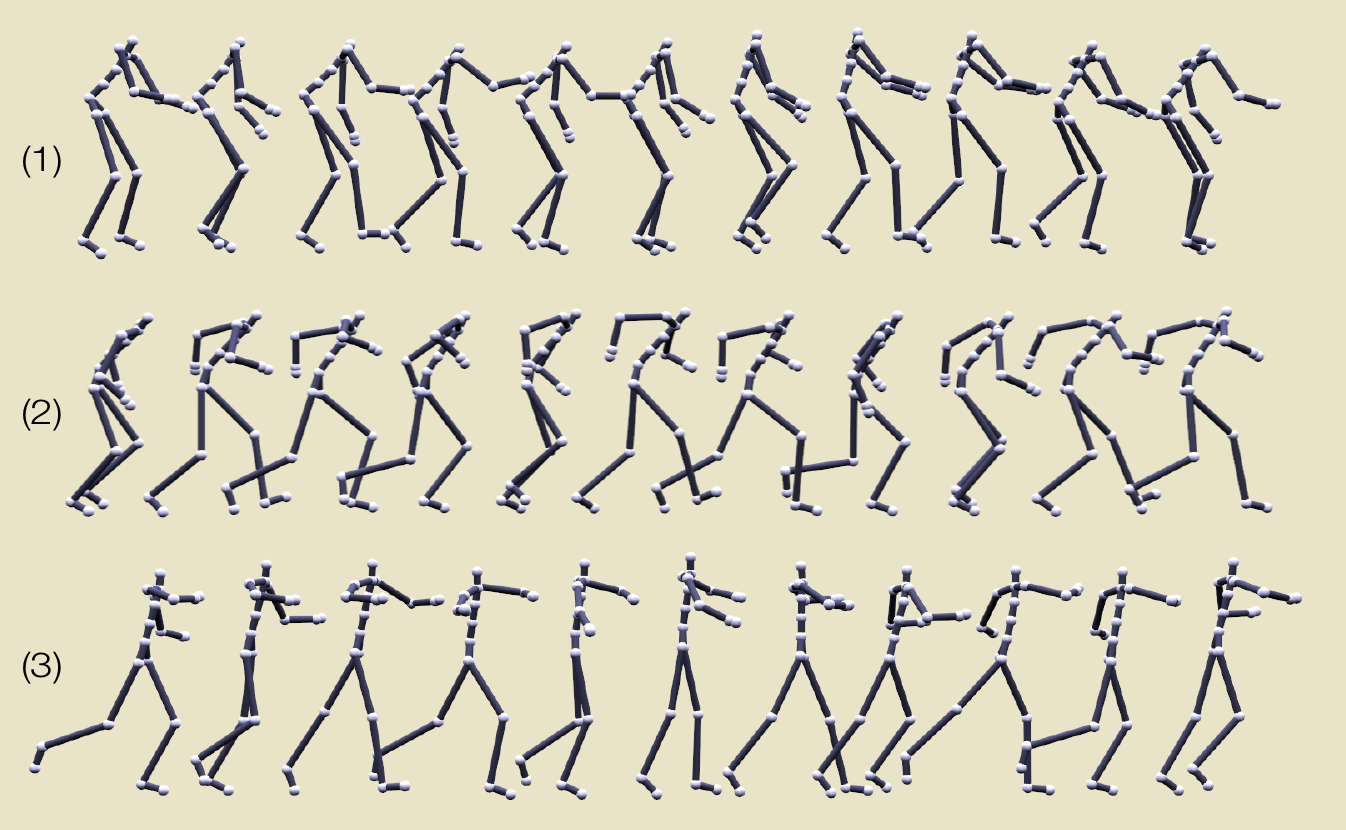}
    \caption{Examples of walk animations in exaggerated styles from existing motion databases \cite{holdenLearningMotionManifolds2015, xiaRealtimeStyleTransfer2015}. (1) "zombie"; (2) "old man"; (3) "childish". Distinguishing each style with no context is not straight-forward.}
    \label{fig:exaggerated_styles}
\end{figure}

We argue that this categorization is too subjective to be considered as an expressive parameter to provide to users. Various factors can introduce bias in style categories, leading edition tools based on them to fail to meet the user's expectation. 

The bias lies in how the data used to calibrate the style categories is recorded. It is usually not captured in the wild, but rather played by actors \cite{xiaRealtimeStyleTransfer2015}. The source sequence are thus often overplayed, exaggerated representations of the actor's and director's vision of the categories (see Fig. \ref{fig:exaggerated_styles}). In the most grounded acquisition procedures, emotions states are elicited in a subject by exposing him to a pre-determined scenario. The bias is however not totally avoided, possibly induced by the design of the experiment \cite{mccambridgeSystematicReviewHawthorne2014, henrichWeirdestPeopleWorld2010}, or even by the divergent interpretations of a common emotion in different cultures \cite{elfenbeinUniversalityCulturalSpecificity2002}. 


The main goal of animation edition research is to provide tools to animators so that they can transform their mental representation of a character motion into tangible animation data. If low-level methods are fine-grained but tedious to manipulate, and if high-level ones are likely biased and take too much of the control out of the user's hands, finding a middle ground is an important open problem. Efficient tooling, allowing easy manipulation for the naïve user and controllable enough for the professional one, are yet to be developed. 

The parameterized edition paradigm, as seen in \cite{rodriguezParameterizedAnimatedActivities2019}, offer interesting perspective in working around the bias problem while still allowing higher-level control. Following this line of thought, this paper proposes an alternative solution, introducing a method to manipulate style from higher-level, but still objective, controls: pose metrics.

\section{Method}

We propose a new character animation edition paradigm in which the animator manipulates intermediate-level parameters called pose metrics, leveraging neural networks to learn and automatically account for complex morphological constraints. We first introduce this new concept, then describe the neural networks pipeline we use in our experiments. We finally describe the method we use to edit whole animation sequences according to a user's input.

In the subsequent sections the term \emph{skeleton} is used to define the virtual armature of a character composed of a hierarchy of joints and bones \cite{magnenat-thalmannJointDependentLocalDeformations1988}. We call \emph{pose} the data vector describing the concatenated position and/or orientation of each joint in the skeleton at a single time frame.  An \emph{animation sequence} is then a sequence of key poses, which represent the character's motion.

In equations the notation is as follow: we denote $A$ an animation sequence containing $N$ frames, with $p = A_n$ the pose at frame $n$. In a skeleton composed of $J$ joints, the pose can be indexed to access a specific joint's data: $p_{head}$.

\subsection{Pose metrics}
 Being able to edit an animation sequence by manipulating higher-level parameters is a desirable capacity, but the literature is polarized between low level controls (operating on crude pose data such as bone position or orientation using forward and inverse kinematics) and high level ones, using \emph{style} labels relating to emotions or behavior (i.e. "angry", "childish", etc. )  \cite{wangAdversarialLearningModeling2020,duStylisticLocomotionModeling2019,abermanUnpairedMotionStyle2020}.
These style controllers lack control in the sense that they produce a stylized animation at once based on a generalized style definition, which users might not share.

As a compromise, filling the gap between these poles, we introduce \emph{pose metrics}: higher-level, objective, style agnostic values, that can be directly computed from a single pose and do not rely on any assumed universal human perceptive behavior.

The vision for these metrics is to be as flexible as possible, so as not to restrict the user. All characters are different, and the concept must make as little assumption as possible on the way they move. That way the user is responsible for designing his own metrics, mentally link them to his own definition of a particular style, and edit the animation through them.

For the sake of an example, imagine an artist editing an animation sequence of a character walking, expressing cold rage. In his mind, the character should be taking short, fast steps, his arms tense, stretched toward the floor. With previous style edition methods, he might be able to hit an "angry" button, maybe use a slider to determine how angry the result should be. Unfortunately, it might be that the system was calibrated with another definition of angry in mind: the character now walks faster, the arms bent and swinging, with the fists close to the torso. Using metrics would allow the user to circumvent this problem. Creating a few metrics such as the angle formed by the elbows, distance between feet, or between hands and the floor, and editing the animation through them, we avoid relying on assumed style characteristics.

The only requirement for a new metric is to be computable from a single pose, and to provide a function to do so. Given a function calculating such a user-provided metric, the proposed associated framework is extendable enough to accommodate it and edit animation sequences accordingly.

\begin{figure}[h!]
    \centering
    \includegraphics[width=0.8\columnwidth]{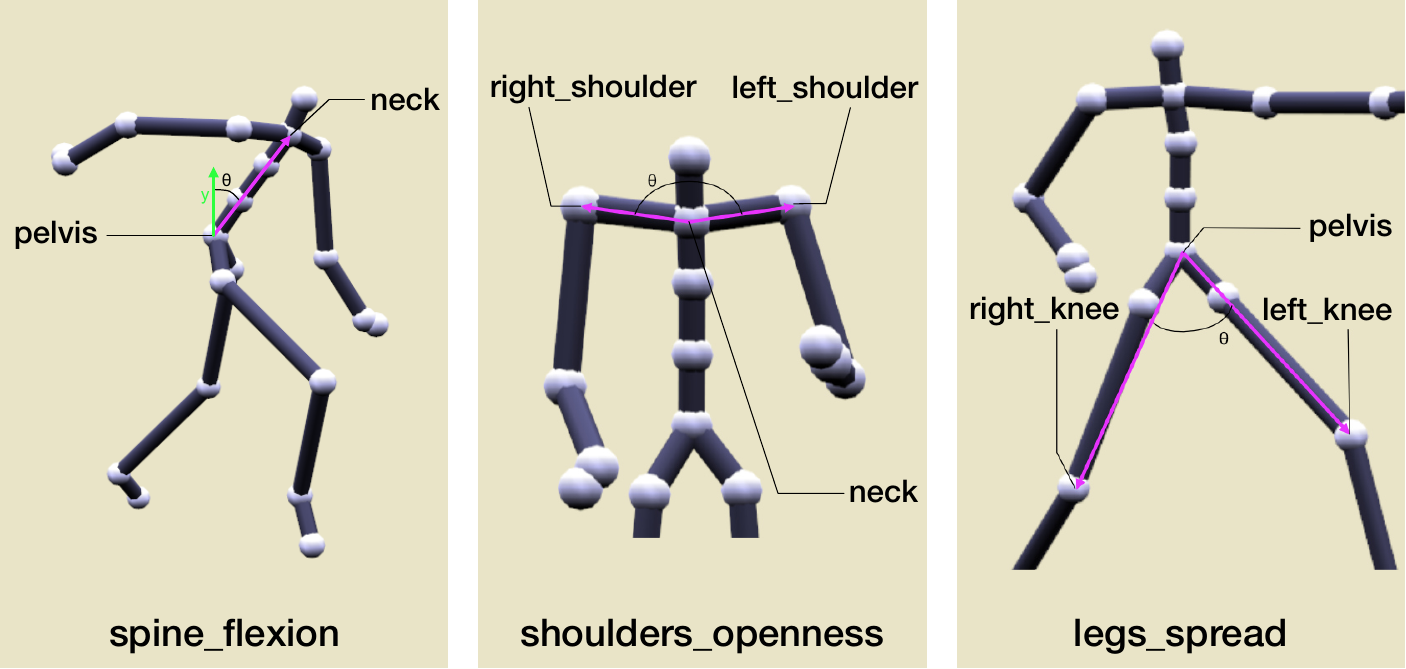}
    \caption{Visual depiction of the three metrics used in this paper's examples.}
    \label{fig:metrics_example}
\end{figure}

\paragraph*{Experimental pose metrics.} As previously explained pose metrics are intended to be provided by the user. In order to demonstrate the application of our system, the following experiments use three sample metrics applicable to a human skeleton: \metric{spine\_flexion}, \metric{shoulders\_openness} and \metric{legs\_spread}.

All three represent the angle formed at the intersection of two vectors $\vec{u}$ and $\vec{v}$ obtained from the pose. Formally the angle at the intersection is computed following Eq. \ref{eq:vector_angle}.

\begin{equation}
    \label{eq:vector_angle}
    \theta(\vec{u}, \Vec{v}) = \arccos \frac{
    \Vec{u} \cdot \Vec{v}
    }
    {
        \lVert \Vec{u} \rVert \cdot \lVert \Vec{v} \rVert
    }
\end{equation}
The \metric{spine\_flexion} metric represents the angle between $\vec{y}$ the world "up" vector and the spine axis. It is computed as shown in eq. \ref{eq:spine_flexion}. It could be used to bend the character forward when in a hurry, or backward when relaxed.

\begin{equation}
    \label{eq:spine_flexion}
    SF(p) = \theta(p_{neck} - p_{pelvis}, \vec{y})
\end{equation}

Next the metric \metric{shoulders\_openness} (Eq. \ref{eq:shoulders_openness}) describes how open the character's torso/shoulders area is. Example usages might be to raise the character shoulders to denote anxiety or shyness, and to open up to denote a more relaxed or proud behavior.

\begin{equation}
    \label{eq:shoulders_openness}
    SO(p) = \theta(
        p_{spine1} - p_{rshoulder},
        p_{lshoulder} - p_{spine1}
    )
\end{equation}

Finally, \metric{legs\_spread} (Eq.\ref{eq:legs_spread}) denotes how close the character's knees are to each other. Possible usages include restricting walking strides, or tweaking whether the character appears in balance.

\begin{equation}
    \label{eq:legs_spread}
    LS(p) = \theta(
        p_{pelvis} - p_{rknee},
        p_{lknee} - p_{pelvis}
    )
\end{equation}

In this paper's equations, the value $m$ of a metric $M$ for a pose $p$ is denoted as the function $m = M(p)$.

\subsection{Learned metric pipeline}

This section introduces a pipeline through which an animation sequence can be edited using pose metrics. It is designed with modularity in mind, as we want users to be able to decide on which metric should be used at any time, as well as extending it with custom metrics if need be. We first describe the global layout of the pipeline, then each of its pieces separately.
We also explain how the pipeline can be extended by the user, and set up to satisfy any combination of metrics. Finally, we propose a procedure to edit whole animation clips, by taking advantage of our architecture to propagate changes made to a selected frame to the rest of the sequence.

\begin{figure}[h]
    \centering
    \includegraphics[width=0.9\columnwidth]{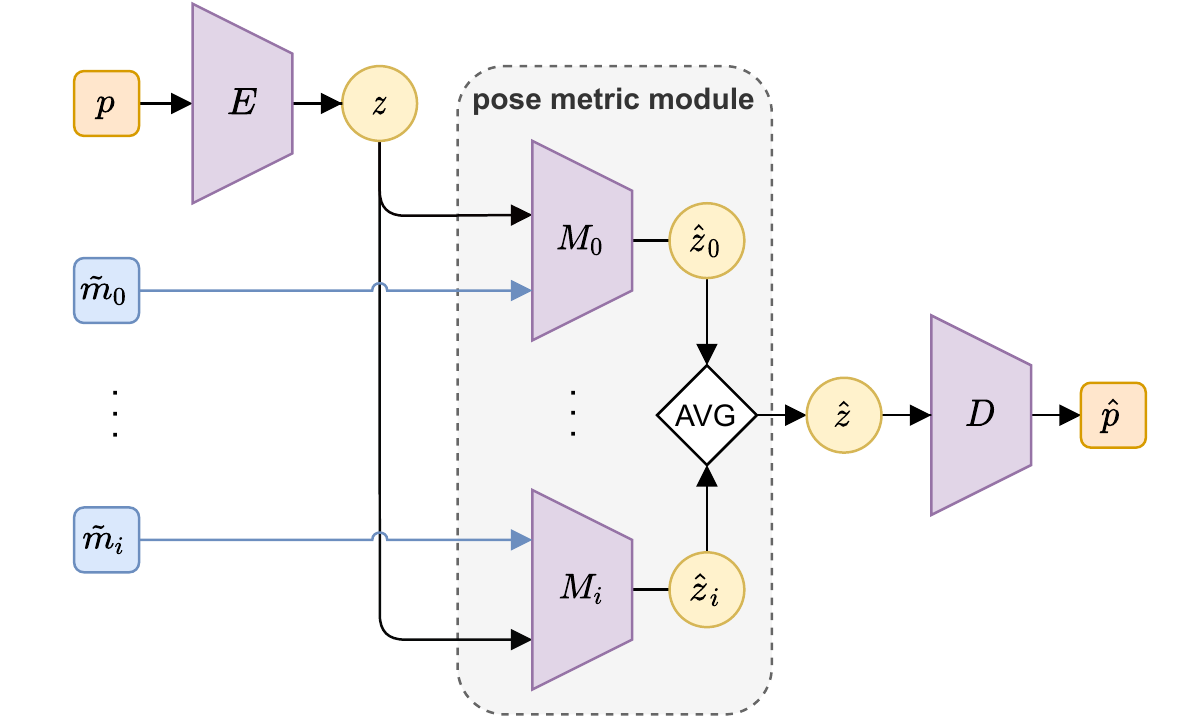}
    \caption{Overview of the proposed pipeline with its pose metric module. The process modifies an input pose $p$ according to an arbitrary number of user-provided target metric values $\tilde{m}_{0...i}$. An encoder-decoder pair ($E$ and $D$) maps poses to and from a latent pose space, on which the other modules operate. Each metric is processed by a dedicated network $M_{0..i}$. Their resulting latent samples $\hat{z}_{0..i}$ are averaged, then mapped back to the pose space.}
    \label{fig:pose_pipeline}
\end{figure}

\paragraph*{Pipeline layout. } 
Our pipeline (Fig. \ref{fig:pose_pipeline}) relies on several small neural networks. We opt for these networks because of their low memory footprint and fast run time. Using smaller independent networks also allows us to easily configure the pipeline by adding, removing or modifying modules.

The outer part of the pipeline is an encoder-decoder model pair whose purpose is to learn a latent pose space, in which a pose representation is more compact and easier to manipulate. As the decoder can only map latent codes to the real pose space, there is no risk of accidentally generating a broken pose when averaging and interpolating samples. The edition itself is done by the metrics network, which operates only on the latent space. An instance of the metric network is trained to modify a latent pose given a \emph{single} metric value. The reduced size of the network and the fact that we re-use the already trained encoder-decoder, mean that training times are fairly short. This in turn further facilitates the possibility for users to design their own metrics, train a network, and incorporate it in the pipeline. 

\paragraph*{Encoder-Decoder.} The encoder $E$ and decoder $D$ networks are trained together to map a pose $p$ to a latent representation $z$, as shown in Eq. \ref{eq:autoencoder}. In practice they are trained to minimize the squared reconstruction error between $p$ and its reconstructed version $\hat{p}$. Both networks are simple 1-layer perceptrons, using 512 units in the hidden layer and rectified linear unit (ReLU) activation functions. In our experiment, the latent size dimension is set to 64.

\begin{equation}
    \label{eq:autoencoder}
    \begin{split}
    z &= E(p) \\
    \hat{p} &= D(z) \\
    \mathcal{L_{\textrm{ED}}} &= \frac{1}{J}\sum_{j=0}^{J}{\big(p_j - \hat{p}_j \big)^2}
    \end{split}
\end{equation}

\paragraph*{Metric network(s)}. A metric network $N_m$ learns to produce a modified latent pose $\hat{z}_m$, taking as input a starting one $z$ concatenated with a target metric value $\targetnotation{m}$ (see Eq. \ref{eq:metric_network_output}).

\begin{equation}
    \label{eq:metric_network_output}
    \hat{z}_m = N_m \big(z \oplus \targetnotation{m}\big)
\end{equation}

To reflect this objective during training, we sample two poses from the same original animation sequence: an input $p = A_t$ and a target $\targetnotation{p} = A_{t+n}$. Sampling from a neighboring frame prevents the target pose from being too far from the starting one and ensures the network learns to \emph{edit} a pose rather than to generate an unrelated one. In our experiments, $n$ is sampled randomly for each iteration in $n \in [-10, 10]$. As we target an interactive edition context, aiming for smaller changes in metric is reasonable.
The input target metric is computed from $\targetnotation{p}$ and fed to the network along with $z = E(p)$. The resulting modified latent pose $\hat{z}$ is mapped back to the pose space using the decoder, i.e. $\hat{p} = D(\hat{z})$. Finally the networks are trained to minimize the mean squared error between the target pose and the modified one, as shown in the loss function in Eq. \ref{eq:metric_loss_fn}.

\begin{equation}
    \label{eq:metric_loss_fn}
    \mathcal{L} = \frac{1}{J} \sum_{j=0}^{J}{\big( \targetnotation{p_j} - \hat{p}_j \big)^2}
\end{equation}

Keeping with the idea of smaller networks we opt for a single layer perceptron with 126 hidden units and ReLU activation. The input layer's number of units is equal to the size of the auto-encoder's latent dimension plus one (to account for the metric value). The output is the same size as the latent dimension. We empirically find that this architecture suits our needs, larger/deeper networks yielding no perceptible value. If a users deemed a different one more suitable for one of his metrics, the only requirements would be on the size of the input and output layers.

\paragraph*{Data}. We use a pose dataset of about 1.5 million poses as described by Victor \etal \cite{victorLearningbasedPoseEdition2021}. The dataset is obtained by merging several pre-existing animation databases\cite{fouratiEmilyaEmotionalBody2014, carnegiemellonuniversityCMUGraphicsLab2003, holdenDeepLearningFramework2016}.
Each animation clip is retargeted to a common skeleton \cite{holdenLearningMotionManifolds2015}. The poses data is composed of the positions of each joint relative to the projection of the pelvis joint on the floor. When used to train the networks, the dataset is normalized by subtracting the mean and dividing by the standard deviation of each component value.

\paragraph*{Training parameters}. The encoder-decoder and metrics networks are trained separately, but using the same hyper-parameters: using the Adam optimizer \cite{kingmaAdamMethodStochastic2015} with a learning rate of $0.0001$, and batches of 1024 pose pairs until convergence, which takes around 15 minutes on an NVIDIA Quadro T1000.

\paragraph*{Pipeline modularity} 
In cases where it is desirable to edit multiple metrics at once, it is possible to combine the results of a set of metric networks. As the output of the network is a latent representation of a pose we can combine them. In \cite{victorLearningbasedPoseEdition2021}, the authors ran multiple networks sequentially. We instead suggest averaging multiple latent poses, so that we don't have to pick an order for the modules.

As explicited in Eq. \ref{eq:metrics_average}, the source pose $p$ is first mapped to the latent space, then each specialized metric network $M{0..i}$ produces modified latent result. These intermediary results are all averaged, and finally mapped back to pose space by the decoder.

\begin{equation}
    \label{eq:metrics_average}
    \hat{p} = D\Big(\frac{1}{n}\sum_{i=0}^{n}{M_i\big(E(p)\big)}\Big) 
\end{equation}

This method operating on single key-frames, can be used conjointly with other posing solutions, and integrated with interpolation methods, to offer as much control to the user as possible. Using a shared latent space also allow us to incorporate other modules from the literature into the pipeline. For example, the IK solver from \cite{victorLearningbasedPoseEdition2021} operates on a similar latent space, and, if trained on the pipeline's one, its output can be averaged with the metric network ones. 

\subsection{Animation edition}

The pose pipeline described above only operates on a single pose at once. In a real-world scenario however animators typically edit a specific key frame, expecting their changes to be gradually applied to the neighboring ones. To meet these expectations we propose a method to edit an entire animation sequence through a single set of target metrics. 

\begin{figure}[h!]
    \centering
    \includegraphics[width=0.9\columnwidth]{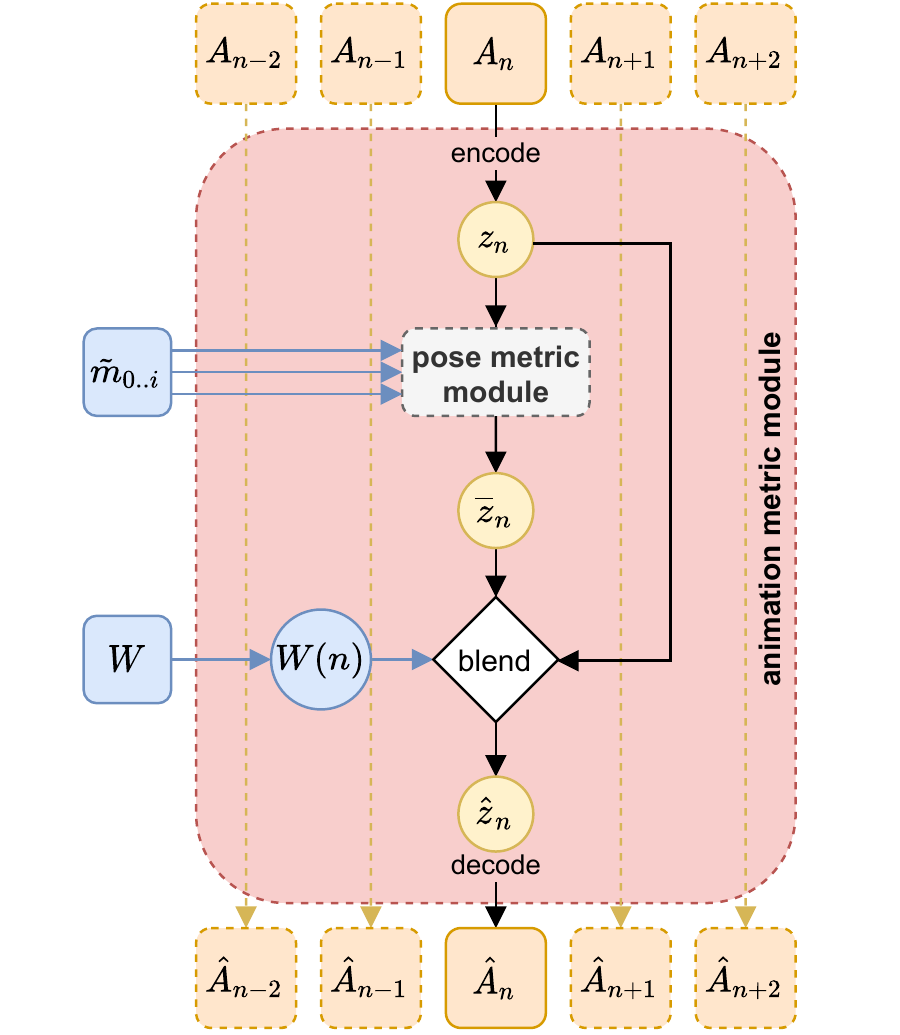}
    \caption{Visual overview of the proposed method to echo modifications made to a key pose $A_n$ of an animation clip. The blend operation occurs between latent poses (source $z_n$ and modified by the metric module $\overline{z}_n$), and use a factor $W(n)$ sampled from the user-provided curve.}
    \label{fig:pipeline_anim}
\end{figure}

Our approach (Fig. \ref{fig:pipeline_anim}) resembles the traditional animation curve paradigm, using a user provided weight curve $W$ to control how much each key pose will be impacted by a modification. We once again leverage the good interpolation properties of the latent space. 

To edit an animation $A$, the user selects a key frame $A_n$ and provides an arbitrary number of target metrics values $\tilde{m}_{0..i}$ as well as a weight curve $W$. For each frame $A_t$ in the sequence, a latent representation $z_t$ is encoded. Each of them is then processed by the pose metric module, resulting in an intermediary latent pose $\overline{z}_t$. The final modified latent pose $\hat{z}_t$ is given by blending the between $z_t$ and $\overline{z}_t$ by the $W(t)$ factor (Eq. \ref{eq:interpolation}).

\begin{equation}
    \label{eq:interpolation}
    \hat{z}_t = \big(1 - W(t)\big) \cdot z_t + W(t) \cdot \overline{z}_t
\end{equation}

In our experiments $W$ is a hat function peaking at $1$ at the selected frame, and reaching $0$ at the clips extremities. This weight function is however an external parameter: experiments with a sinusoidal yield comparable results, and the curve could be edited by animators to fit their expectations, much like existing animation curves. We also note that each one of the clip's poses is processed independently from the others. The edition can thus happen in parallel, maintaining real time interactivity.

\section{Results}

This section presents visual animation edition results of the pipeline in different scenarios. More examples can be found in the accompanying video.

\subsection{Pose edition}

\begin{figure}[h!]
    \centering
    \includegraphics[width=0.9\columnwidth]{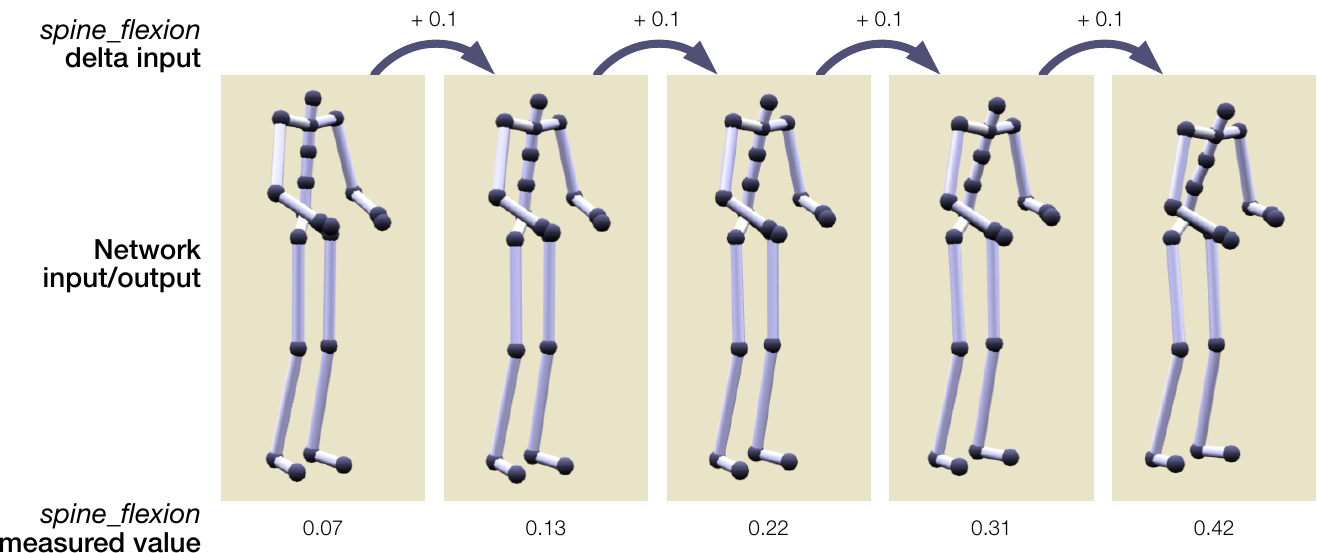}
    \caption{Incremental modification of a pose by providing a target $spine\_flexion$ input. On each step, the user's input is a $0.1$ increase over the computed metric value.}
    \label{fig:spine_flexion_pose}
\end{figure}

In Fig. \ref{fig:spine_flexion_pose}, a single pose is modified by a user by providing gradually increasing target \metric{spine\_flexion} metric values. The results respect the desired changes as shown by the measured metric values below each pose. While it could be argued that changing the orientation angle of the pelvis bone would produce similar results, a closer look to the resulting poses illustrates the interesting properties of the latent pose space. At every iteration, the skeleton's arms are slightly raised, and its knees bent, so as to compensate the loss of balance induced by the flexion of the spine. 

\subsection{Modular pipeline}

\begin{figure}[h]
    \centering
    \includegraphics[width=0.6\columnwidth]{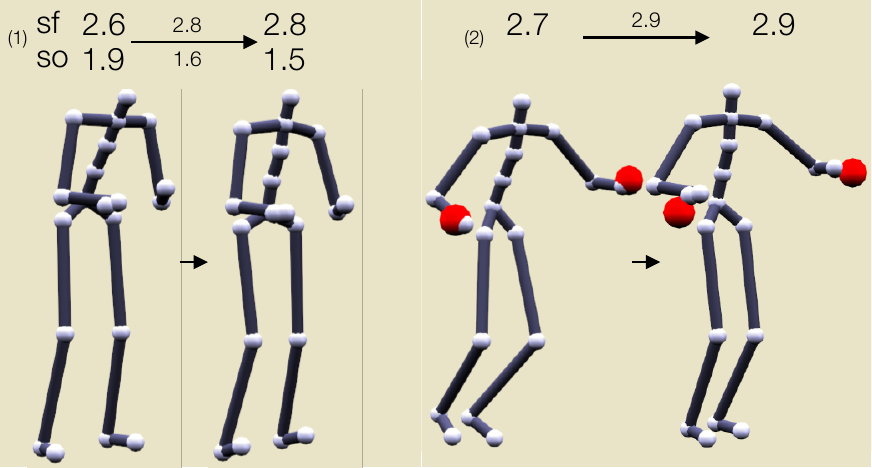}
    \caption{Sample results of the pipeline using multiple modules. In (1), \metric{shoulders\_openness} (so) and \metric{spine\_flexion} are used simultaneously. In (2) we incorporate the latent inverse kinematic solver from \cite{victorLearningbasedPoseEdition2021} into the pipeline, using it along \metric{spine\_flexion}.}
    \label{fig:multiple_metrics_results}
\end{figure}

Fig. \ref{fig:multiple_metrics_results} presents some results obtained by averaging the latent outputs from multiple modules. The first example show the result of modifying the \metric{spine\_flexion} and \metric{shoulders\_openness} metrics at the same time. Both modifications can be seen on the skeleton: its spine is straightened, and its shoulders lowered. The rest of the pose is also slightly edited to accommodate for those change: the hands are lowered, and knees bent a little further.

The second one shows the result of using the \metric{spine\_flexion} metric module along with the neural IK method from \cite{victorLearningbasedPoseEdition2021}. Averaging the latent outputs leads our method to find a compromise between each provided constraint. Here, the skeleton's back is straightened, and the IK targets for the hands are reached for.

\subsection{Animation edition}

\begin{figure}[b]
    \centering
    \includegraphics[width=0.9\columnwidth]{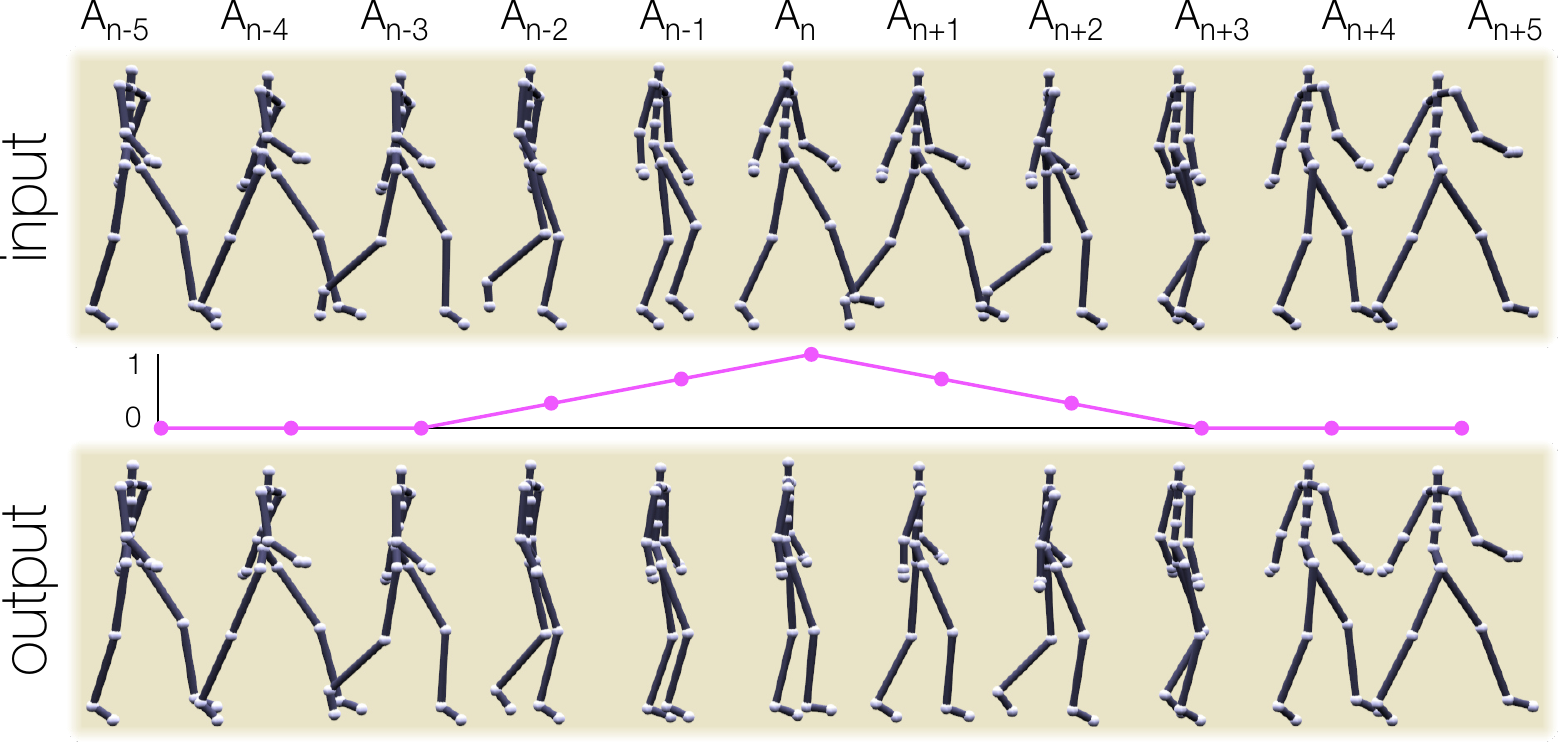}
    \caption{Editing an animation clip according to the user input. The output is produced by the pipeline given a selected frame $A_n$, a target value for the \metric{legs\_spread} metric, and a weight curve. }
    \label{fig:animation_edition_result}
\end{figure}

Fig. \ref{fig:animation_edition_result} showcases the result of an animation $A$ being edited using the pipeline described earlier. The user selects a frame $A_n$, provides a target \metric{leg\_spread} metric value, and and interpolation factor curve, a hat function peaking at one on the $A_n$ and reaching down to 0 on both sides at $A_{n-3}$ and $A_{n+3}$. The remaining frames use a factor of 0.

Each frame pose is processed separately by the network, and the result is given by interpolating the latent representations of the input and output poses, using the frame-specific factor. 

As a result, the output animation $\hat{A}$ is modified, with the impact being most important on the selected frame and gradually lowering on each side. At frame $n$ the output pose is modified the most, and the reduced legs angle is very visible. At frames $n-2$ and $n+2$ this impact is barely visible, while further frames are not modified at all, as expected.

Interpolating between original and modified poses allow the whole sequence to keep its temporal coherence even though the pipeline processes one frame at a time, with no knowledge of the past and future frame context. We also take advantage of the properties of latent spaces: any latent interpolation result is mapped back to the real pose space, thus avoiding artifacts such as skeleton interpenetration.

\section{Conclusion}
In order to avoid the pitfalls of categorizing animation styles, we introduce the concept of pose metrics, values computed on a single pose that user can define themselves to fit their \emph{own} perception of styles. We propose a pipeline to edit an animation by manipulating these metrics, which uses a modular set of tiny neural networks, running in real time. A user can extend the pipeline by designing a custom metric and training an associated metric network in a short 15 minutes.

The main shortcoming of this approach is that it is limited to a single standardized humanoid skeleton, with which the networks are trained. This is a common problem with data-driven methods: in order to adapt to a new skeleton, the network would need to be retrained. This could be difficult if access to existing animation data for the new skeleton is limited, but recent progress in animation retargeting \cite{abermanSkeletonawareNetworksDeep2020} could help by allowing more accessible data augmentation. The modularity of metrics helps here as well, as different skeleton will require a new set of metrics.

Future work will focus on grounding the proposed pipeline: securing the process by automatically detecting overlapping metrics, and conducting user studies to evaluate the usability of the method.

\bibliographystyle{ACM-Reference-Format}
\bibliography{bib}


\begin{thebibliography}{39}


\ifx \showCODEN    \undefined \def \showCODEN     #1{\unskip}     \fi
\ifx \showDOI      \undefined \def \showDOI       #1{#1}\fi
\ifx \showISBNx    \undefined \def \showISBNx     #1{\unskip}     \fi
\ifx \showISBNxiii \undefined \def \showISBNxiii  #1{\unskip}     \fi
\ifx \showISSN     \undefined \def \showISSN      #1{\unskip}     \fi
\ifx \showLCCN     \undefined \def \showLCCN      #1{\unskip}     \fi
\ifx \shownote     \undefined \def \shownote      #1{#1}          \fi
\ifx \showarticletitle \undefined \def \showarticletitle #1{#1}   \fi
\ifx \showURL      \undefined \def \showURL       {\relax}        \fi
\providecommand\bibfield[2]{#2}
\providecommand\bibinfo[2]{#2}
\providecommand\natexlab[1]{#1}
\providecommand\showeprint[2][]{arXiv:#2}

\bibitem[Aberman et~al\mbox{.}(2020a)]%
        {abermanSkeletonawareNetworksDeep2020}
\bibfield{author}{\bibinfo{person}{Kfir Aberman}, \bibinfo{person}{Peizhuo Li},
  \bibinfo{person}{Dani Lischinski}, \bibinfo{person}{Olga {Sorkine-Hornung}},
  \bibinfo{person}{Daniel {Cohen-Or}}, {and} \bibinfo{person}{Baoquan Chen}.}
  \bibinfo{year}{2020}\natexlab{a}.
\newblock \showarticletitle{Skeleton-Aware Networks for Deep Motion
  Retargeting}.
\newblock \bibinfo{journal}{\emph{ACM Transactions on Graphics}}
  \bibinfo{volume}{39}, \bibinfo{number}{4} (\bibinfo{date}{July}
  \bibinfo{year}{2020}), \bibinfo{pages}{62:62:1--62:62:14}.
\newblock
\showISSN{0730-0301}
\urldef\tempurl%
\url{https://doi.org/10.1145/3386569.3392462}
\showDOI{\tempurl}


\bibitem[Aberman et~al\mbox{.}(2020b)]%
        {abermanUnpairedMotionStyle2020}
\bibfield{author}{\bibinfo{person}{Kfir Aberman}, \bibinfo{person}{Yijia Weng},
  \bibinfo{person}{Dani Lischinski}, \bibinfo{person}{Daniel {Cohen-Or}}, {and}
  \bibinfo{person}{Baoquan Chen}.} \bibinfo{year}{2020}\natexlab{b}.
\newblock \showarticletitle{Unpaired Motion Style Transfer from Video to
  Animation}.
\newblock \bibinfo{journal}{\emph{ACM Transactions on Graphics}}
  \bibinfo{volume}{39}, \bibinfo{number}{4} (\bibinfo{date}{July}
  \bibinfo{year}{2020}), \bibinfo{pages}{64:64:1--64:64:12}.
\newblock
\showISSN{0730-0301}
\urldef\tempurl%
\url{https://doi.org/10.1145/3386569.3392469}
\showDOI{\tempurl}


\bibitem[Aristidou and Lasenby(2011)]%
        {aristidouFABRIKFastIterative2011}
\bibfield{author}{\bibinfo{person}{Andreas Aristidou} {and}
  \bibinfo{person}{Joan Lasenby}.} \bibinfo{year}{2011}\natexlab{}.
\newblock \showarticletitle{{{FABRIK}}: {{A}} Fast, Iterative Solver for the
  {{Inverse Kinematics}} Problem}.
\newblock \bibinfo{journal}{\emph{Graphical Models}} \bibinfo{volume}{73},
  \bibinfo{number}{5} (\bibinfo{date}{Sept.} \bibinfo{year}{2011}),
  \bibinfo{pages}{243--260}.
\newblock
\showISSN{1524-0703}
\urldef\tempurl%
\url{https://doi.org/10.1016/j.gmod.2011.05.003}
\showDOI{\tempurl}


\bibitem[Aristidou et~al\mbox{.}(2017)]%
        {aristidouEmotionControlUnstructured2017}
\bibfield{author}{\bibinfo{person}{Andreas Aristidou}, \bibinfo{person}{Qiong
  Zeng}, \bibinfo{person}{Efstathios Stavrakis}, \bibinfo{person}{KangKang
  Yin}, \bibinfo{person}{Daniel {Cohen-Or}}, \bibinfo{person}{Yiorgos
  Chrysanthou}, {and} \bibinfo{person}{Baoquan Chen}.}
  \bibinfo{year}{2017}\natexlab{}.
\newblock \showarticletitle{Emotion Control of Unstructured Dance Movements}.
  In \bibinfo{booktitle}{\emph{Proceedings of the {{ACM SIGGRAPH}} /
  {{Eurographics Symposium}} on {{Computer Animation}}}}
  \emph{(\bibinfo{series}{{{SCA}} '17})}. \bibinfo{publisher}{{Association for
  Computing Machinery}}, \bibinfo{address}{{New York, NY, USA}},
  \bibinfo{pages}{1--10}.
\newblock
\showISBNx{978-1-4503-5091-4}
\urldef\tempurl%
\url{https://doi.org/10.1145/3099564.3099566}
\showDOI{\tempurl}


\bibitem[Brand and Hertzmann(2000)]%
        {brandStyleMachines2000}
\bibfield{author}{\bibinfo{person}{Matthew Brand} {and} \bibinfo{person}{Aaron
  Hertzmann}.} \bibinfo{year}{2000}\natexlab{}.
\newblock \showarticletitle{Style {{Machines}}}. In
  \bibinfo{booktitle}{\emph{Proceedings of the 27th {{Annual Conference}} on
  {{Computer Graphics}} and {{Interactive Techniques}}}}
  \emph{(\bibinfo{series}{{{SIGGRAPH}} '00})}. \bibinfo{publisher}{{ACM
  Press/Addison-Wesley Publishing Co.}}, \bibinfo{address}{{New York, NY,
  USA}}, \bibinfo{pages}{183--192}.
\newblock
\showISBNx{978-1-58113-208-3}
\urldef\tempurl%
\url{https://doi.org/10.1145/344779.344865}
\showDOI{\tempurl}


\bibitem[Catmull(1978)]%
        {catmullProblemsComputerassistedAnimation1978}
\bibfield{author}{\bibinfo{person}{Edwin Catmull}.}
  \bibinfo{year}{1978}\natexlab{}.
\newblock \showarticletitle{The {{Problems}} of {{Computer-assisted
  Animation}}}. In \bibinfo{booktitle}{\emph{Proceedings of the 5th {{Annual
  Conference}} on {{Computer Graphics}} and {{Interactive Techniques}}}}
  \emph{(\bibinfo{series}{{{SIGGRAPH}} '78})}. \bibinfo{publisher}{{ACM}},
  \bibinfo{address}{{New York, NY, USA}}, \bibinfo{pages}{348--353}.
\newblock
\urldef\tempurl%
\url{https://doi.org/10.1145/800248.807414}
\showDOI{\tempurl}


\bibitem[Catmull and Rom(1974)]%
        {catmullClassLocalInterpolating1974}
\bibfield{author}{\bibinfo{person}{Edwin Catmull} {and}
  \bibinfo{person}{Raphael Rom}.} \bibinfo{year}{1974}\natexlab{}.
\newblock \showarticletitle{A Class of Local Interpolating Splines}.
\newblock In \bibinfo{booktitle}{\emph{Computer {{Aided Geometric Design}}}},
  \bibfield{editor}{\bibinfo{person}{ROBERT~E. Barnhill} {and}
  \bibinfo{person}{RICHARD~F. Riesenfeld}} (Eds.).
  \bibinfo{publisher}{{Academic Press}}, \bibinfo{pages}{317--326}.
\newblock
\showISBNx{978-0-12-079050-0}
\urldef\tempurl%
\url{https://doi.org/10.1016/B978-0-12-079050-0.50020-5}
\showDOI{\tempurl}


\bibitem[Chai and Hodgins(2007)]%
        {chaiConstraintbasedMotionOptimization2007}
\bibfield{author}{\bibinfo{person}{Jinxiang Chai} {and}
  \bibinfo{person}{Jessica~K. Hodgins}.} \bibinfo{year}{2007}\natexlab{}.
\newblock \showarticletitle{Constraint-Based Motion Optimization Using a
  Statistical Dynamic Model}. In \bibinfo{booktitle}{\emph{{{ACM SIGGRAPH}}
  2007 Papers}} \emph{(\bibinfo{series}{{{SIGGRAPH}} '07})}.
  \bibinfo{publisher}{{Association for Computing Machinery}},
  \bibinfo{address}{{New York, NY, USA}}, \bibinfo{pages}{8--es}.
\newblock
\showISBNx{978-1-4503-7836-9}
\urldef\tempurl%
\url{https://doi.org/10.1145/1275808.1276387}
\showDOI{\tempurl}


\bibitem[Choi et~al\mbox{.}(2016)]%
        {choiSketchiMoSketchbasedMotion2016}
\bibfield{author}{\bibinfo{person}{Byungkuk Choi}, \bibinfo{person}{Roger
  {Blanco i Ribera}}, \bibinfo{person}{J.~P. Lewis}, \bibinfo{person}{Yeongho
  Seol}, \bibinfo{person}{Seokpyo Hong}, \bibinfo{person}{Haegwang Eom},
  \bibinfo{person}{Sunjin Jung}, {and} \bibinfo{person}{Junyong Noh}.}
  \bibinfo{year}{2016}\natexlab{}.
\newblock \showarticletitle{{{SketchiMo}}: {{Sketch-based Motion Editing}} for
  {{Articulated Characters}}}.
\newblock \bibinfo{journal}{\emph{ACM Trans. Graph.}} \bibinfo{volume}{35},
  \bibinfo{number}{4} (\bibinfo{date}{July} \bibinfo{year}{2016}),
  \bibinfo{pages}{146:1--146:12}.
\newblock
\showISSN{0730-0301}
\urldef\tempurl%
\url{https://doi.org/10.1145/2897824.2925970}
\showDOI{\tempurl}


\bibitem[Ciccone et~al\mbox{.}(2019)]%
        {cicconeTangentspaceOptimizationInteractive2019}
\bibfield{author}{\bibinfo{person}{Lo{\"i}c Ciccone}, \bibinfo{person}{Cengiz
  {\"O}ztireli}, {and} \bibinfo{person}{Robert~W. Sumner}.}
  \bibinfo{year}{2019}\natexlab{}.
\newblock \showarticletitle{Tangent-Space Optimization for Interactive
  Animation Control}.
\newblock \bibinfo{journal}{\emph{ACM Transactions on Graphics}}
  \bibinfo{volume}{38}, \bibinfo{number}{4} (\bibinfo{date}{July}
  \bibinfo{year}{2019}), \bibinfo{pages}{1--10}.
\newblock
\showISSN{07300301}
\urldef\tempurl%
\url{https://doi.org/10.1145/3306346.3322938}
\showDOI{\tempurl}


\bibitem[Du et~al\mbox{.}(2019)]%
        {duStylisticLocomotionModeling2019}
\bibfield{author}{\bibinfo{person}{Han Du}, \bibinfo{person}{Erik Herrmann},
  \bibinfo{person}{Janis Sprenger}, \bibinfo{person}{Klaus Fischer}, {and}
  \bibinfo{person}{Philipp Slusallek}.} \bibinfo{year}{2019}\natexlab{}.
\newblock \showarticletitle{Stylistic {{Locomotion Modeling}} and {{Synthesis
  Using Variational Generative Models}}}. In \bibinfo{booktitle}{\emph{Motion,
  {{Interaction}} and {{Games}}}} \emph{(\bibinfo{series}{{{MIG}} '19})}.
  \bibinfo{publisher}{{ACM}}, \bibinfo{address}{{New York, NY, USA}},
  \bibinfo{pages}{32:1--32:10}.
\newblock
\showISBNx{978-1-4503-6994-7}
\urldef\tempurl%
\url{https://doi.org/10.1145/3359566.3360083}
\showDOI{\tempurl}


\bibitem[Durupinar et~al\mbox{.}(2016)]%
        {durupinarPERFORMPerceptualApproach2016}
\bibfield{author}{\bibinfo{person}{Funda Durupinar}, \bibinfo{person}{Mubbasir
  Kapadia}, \bibinfo{person}{Susan Deutsch}, \bibinfo{person}{Michael Neff},
  {and} \bibinfo{person}{Norman~I. Badler}.} \bibinfo{year}{2016}\natexlab{}.
\newblock \showarticletitle{{{PERFORM}}: {{Perceptual Approach}} for {{Adding
  OCEAN Personality}} to {{Human Motion Using Laban Movement Analysis}}}.
\newblock \bibinfo{journal}{\emph{ACM Trans. Graph.}} \bibinfo{volume}{36},
  \bibinfo{number}{1} (\bibinfo{date}{Oct.} \bibinfo{year}{2016}).
\newblock
\showISSN{0730-0301}
\urldef\tempurl%
\url{https://doi.org/10.1145/2983620}
\showDOI{\tempurl}


\bibitem[Ekman and Friesen(1967)]%
        {ekmanHeadBodyCues1967}
\bibfield{author}{\bibinfo{person}{P. Ekman} {and} \bibinfo{person}{W.~V.
  Friesen}.} \bibinfo{year}{1967}\natexlab{}.
\newblock \showarticletitle{Head and Body Cues in the Judgment of Emotion: A
  Reformulation}.
\newblock \bibinfo{journal}{\emph{Perceptual and motor skills}}
  \bibinfo{volume}{24}, \bibinfo{number}{3} (\bibinfo{date}{June}
  \bibinfo{year}{1967}), \bibinfo{pages}{711--724}.
\newblock
\showISSN{1558-688X}
\urldef\tempurl%
\url{https://doi.org/10.2466/pms.1967.24.3.711}
\showDOI{\tempurl}


\bibitem[Elfenbein and Ambady(2002)]%
        {elfenbeinUniversalityCulturalSpecificity2002}
\bibfield{author}{\bibinfo{person}{Hillary~Anger Elfenbein} {and}
  \bibinfo{person}{Nalini Ambady}.} \bibinfo{year}{2002}\natexlab{}.
\newblock \showarticletitle{On the Universality and Cultural Specificity of
  Emotion Recognition: {{A}} Meta-Analysis}.
\newblock \bibinfo{journal}{\emph{Psychological Bulletin}}
  \bibinfo{volume}{128}, \bibinfo{number}{2} (\bibinfo{year}{2002}),
  \bibinfo{pages}{203--235}.
\newblock
\showISSN{1939-1455}
\urldef\tempurl%
\url{https://doi.org/10.1037/0033-2909.128.2.203}
\showDOI{\tempurl}


\bibitem[Fourati and Pelachaud(2014)]%
        {fouratiEmilyaEmotionalBody2014}
\bibfield{author}{\bibinfo{person}{Nesrine Fourati} {and}
  \bibinfo{person}{Catherine Pelachaud}.} \bibinfo{year}{2014}\natexlab{}.
\newblock \showarticletitle{Emilya: {{Emotional}} Body Expression in Daily
  Actions Database}. In \bibinfo{booktitle}{\emph{Proceedings of the {{Ninth
  International Conference}} on {{Language Resources}} and {{Evaluation}}
  ({{LREC-2014}})}}. \bibinfo{publisher}{{European Languages Resources
  Association (ELRA)}}, \bibinfo{address}{{Reykjavik, Iceland}},
  \bibinfo{pages}{3486--3493}.
\newblock


\bibitem[Garcia et~al\mbox{.}(2019)]%
        {garciaSpatialMotionDoodles2019}
\bibfield{author}{\bibinfo{person}{Maxime Garcia}, \bibinfo{person}{Remi
  Ronfard}, {and} \bibinfo{person}{Marie-Paule Cani}.}
  \bibinfo{year}{2019}\natexlab{}.
\newblock \showarticletitle{Spatial {{Motion Doodles}}: {{Sketching Animation}}
  in {{VR Using Hand Gestures}} and {{Laban Motion Analysis}}}. In
  \bibinfo{booktitle}{\emph{Motion, {{Interaction}} and {{Games}}}}
  \emph{(\bibinfo{series}{{{MIG}} '19})}. \bibinfo{publisher}{{Association for
  Computing Machinery}}, \bibinfo{address}{{New York, NY, USA}},
  \bibinfo{pages}{1--10}.
\newblock
\showISBNx{978-1-4503-6994-7}
\urldef\tempurl%
\url{https://doi.org/10.1145/3359566.3360061}
\showDOI{\tempurl}


\bibitem[Guay et~al\mbox{.}(2013)]%
        {guayLineActionIntuitive2013}
\bibfield{author}{\bibinfo{person}{Martin Guay}, \bibinfo{person}{Marie-Paule
  Cani}, {and} \bibinfo{person}{R{\'e}mi Ronfard}.}
  \bibinfo{year}{2013}\natexlab{}.
\newblock \showarticletitle{The Line of Action: An Intuitive Interface for
  Expressive Character Posing}.
\newblock \bibinfo{journal}{\emph{ACM Transactions on Graphics}}
  \bibinfo{volume}{32}, \bibinfo{number}{6} (\bibinfo{date}{Nov.}
  \bibinfo{year}{2013}), \bibinfo{pages}{1--8}.
\newblock
\showISSN{0730-0301, 1557-7368}
\urldef\tempurl%
\url{https://doi.org/10.1145/2508363.2508397}
\showDOI{\tempurl}


\bibitem[Harvey et~al\mbox{.}(2020)]%
        {harveyRobustMotionInbetweening2020}
\bibfield{author}{\bibinfo{person}{F{\'e}lix~G. Harvey}, \bibinfo{person}{Mike
  Yurick}, \bibinfo{person}{Derek Nowrouzezahrai}, {and}
  \bibinfo{person}{Christopher Pal}.} \bibinfo{year}{2020}\natexlab{}.
\newblock \showarticletitle{Robust Motion In-Betweening}.
\newblock \bibinfo{journal}{\emph{ACM Transactions on Graphics}}
  \bibinfo{volume}{39}, \bibinfo{number}{4} (\bibinfo{date}{July}
  \bibinfo{year}{2020}), \bibinfo{pages}{60:60:1--60:60:12}.
\newblock
\showISSN{0730-0301}
\urldef\tempurl%
\url{https://doi.org/10.1145/3386569.3392480}
\showDOI{\tempurl}


\bibitem[Henrich et~al\mbox{.}(2010)]%
        {henrichWeirdestPeopleWorld2010}
\bibfield{author}{\bibinfo{person}{Joseph Henrich}, \bibinfo{person}{Steven~J.
  Heine}, {and} \bibinfo{person}{Ara Norenzayan}.}
  \bibinfo{year}{2010}\natexlab{}.
\newblock \showarticletitle{The Weirdest People in the World?}
\newblock \bibinfo{journal}{\emph{Behavioral and Brain Sciences}}
  \bibinfo{volume}{33}, \bibinfo{number}{2-3} (\bibinfo{date}{June}
  \bibinfo{year}{2010}), \bibinfo{pages}{61--83}.
\newblock
\showISSN{1469-1825, 0140-525X}
\urldef\tempurl%
\url{https://doi.org/10.1017/S0140525X0999152X}
\showDOI{\tempurl}


\bibitem[Holden et~al\mbox{.}(2016)]%
        {holdenDeepLearningFramework2016}
\bibfield{author}{\bibinfo{person}{Daniel Holden}, \bibinfo{person}{Jun Saito},
  {and} \bibinfo{person}{Taku Komura}.} \bibinfo{year}{2016}\natexlab{}.
\newblock \showarticletitle{A Deep Learning Framework for Character Motion
  Synthesis and Editing}.
\newblock \bibinfo{journal}{\emph{ACM Transactions on Graphics}}
  \bibinfo{volume}{35}, \bibinfo{number}{4} (\bibinfo{date}{July}
  \bibinfo{year}{2016}), \bibinfo{pages}{1--11}.
\newblock
\showISSN{07300301}
\urldef\tempurl%
\url{https://doi.org/10.1145/2897824.2925975}
\showDOI{\tempurl}


\bibitem[Holden et~al\mbox{.}(2015)]%
        {holdenLearningMotionManifolds2015}
\bibfield{author}{\bibinfo{person}{Daniel Holden}, \bibinfo{person}{Jun Saito},
  \bibinfo{person}{Taku Komura}, {and} \bibinfo{person}{Thomas Joyce}.}
  \bibinfo{year}{2015}\natexlab{}.
\newblock \showarticletitle{Learning Motion Manifolds with Convolutional
  Autoencoders}. In \bibinfo{booktitle}{\emph{{{SIGGRAPH ASIA}} 2015
  {{Technical Briefs}} on - {{SA}} '15}}. \bibinfo{publisher}{{ACM Press}},
  \bibinfo{address}{{Kobe, Japan}}, \bibinfo{pages}{1--4}.
\newblock
\showISBNx{978-1-4503-3930-8}
\urldef\tempurl%
\url{https://doi.org/10.1145/2820903.2820918}
\showDOI{\tempurl}


\bibitem[Jacobson et~al\mbox{.}(2014)]%
        {jacobsonTangibleModularInput2014}
\bibfield{author}{\bibinfo{person}{Alec Jacobson}, \bibinfo{person}{Daniele
  Panozzo}, \bibinfo{person}{Oliver Glauser}, \bibinfo{person}{C{\'e}dric
  Pradalier}, \bibinfo{person}{Otmar Hilliges}, {and} \bibinfo{person}{Olga
  {Sorkine-Hornung}}.} \bibinfo{year}{2014}\natexlab{}.
\newblock \showarticletitle{Tangible and Modular Input Device for Character
  Articulation}.
\newblock \bibinfo{journal}{\emph{ACM Transactions on Graphics}}
  \bibinfo{volume}{33}, \bibinfo{number}{4} (\bibinfo{date}{July}
  \bibinfo{year}{2014}), \bibinfo{pages}{82:1--82:12}.
\newblock
\showISSN{0730-0301}
\urldef\tempurl%
\url{https://doi.org/10.1145/2601097.2601112}
\showDOI{\tempurl}


\bibitem[Kingma and Ba(2015)]%
        {kingmaAdamMethodStochastic2015}
\bibfield{author}{\bibinfo{person}{Diederik~P. Kingma} {and}
  \bibinfo{person}{Jimmy Ba}.} \bibinfo{year}{2015}\natexlab{}.
\newblock \showarticletitle{Adam: {{A Method}} for {{Stochastic
  Optimization}}}. In \bibinfo{booktitle}{\emph{{{arXiv}}:1412.6980 [Cs]}}.
  \bibinfo{address}{{San Diego, California, US}}.
\newblock
\urldef\tempurl%
\url{https://doi.org/10.48550/ARXIV.1412.6980}
\showDOI{\tempurl}
\showeprint[arxiv]{1412.6980}~[cs]


\bibitem[Koyama and Goto(2018)]%
        {koyamaOptiMoOptimizationGuidedMotion2018}
\bibfield{author}{\bibinfo{person}{Yuki Koyama} {and} \bibinfo{person}{Masataka
  Goto}.} \bibinfo{year}{2018}\natexlab{}.
\newblock \showarticletitle{{{OptiMo}}: {{Optimization-Guided Motion Editing}}
  for {{Keyframe Character Animation}}}. In
  \bibinfo{booktitle}{\emph{Proceedings of the 2018 {{CHI Conference}} on
  {{Human Factors}} in {{Computing Systems}} - {{CHI}} '18}}.
  \bibinfo{publisher}{{ACM Press}}, \bibinfo{address}{{Montreal QC, Canada}},
  \bibinfo{pages}{1--12}.
\newblock
\showISBNx{978-1-4503-5620-6}
\urldef\tempurl%
\url{https://doi.org/10.1145/3173574.3173735}
\showDOI{\tempurl}


\bibitem[Lasseter(1987)]%
        {lasseterPrinciplesTraditionalAnimation1987}
\bibfield{author}{\bibinfo{person}{John Lasseter}.}
  \bibinfo{year}{1987}\natexlab{}.
\newblock \showarticletitle{Principles of Traditional Animation Applied to
  {{3D}} Computer Animation}.
\newblock \bibinfo{journal}{\emph{ACM SIGGRAPH Computer Graphics}}
  \bibinfo{volume}{21}, \bibinfo{number}{4} (\bibinfo{date}{Aug.}
  \bibinfo{year}{1987}), \bibinfo{pages}{35--44}.
\newblock
\showISSN{0097-8930}
\urldef\tempurl%
\url{https://doi.org/10.1145/37402.37407}
\showDOI{\tempurl}


\bibitem[{Magnenat-Thalmann} et~al\mbox{.}(1988)]%
        {magnenat-thalmannJointDependentLocalDeformations1988}
\bibfield{author}{\bibinfo{person}{Nadia {Magnenat-Thalmann}},
  \bibinfo{person}{Richard Laperrire}, {and} \bibinfo{person}{Daniel
  Thalmann}.} \bibinfo{year}{1988}\natexlab{}.
\newblock \showarticletitle{Joint-{{Dependent Local Deformations}} for {{Hand
  Animation}} and {{Object Grasping}}}. In
  \bibinfo{booktitle}{\emph{Proceedings on {{Graphics Interface}} '88}}
  \emph{(\bibinfo{series}{{{GI}} '88})}. \bibinfo{publisher}{{Canadian
  Man-Computer Communications Society}}, \bibinfo{address}{{Edmonton, Alberta,
  Canada}}, \bibinfo{pages}{26--33}.
\newblock
\urldef\tempurl%
\url{https://doi.org/10.20380/gi1988.04}
\showDOI{\tempurl}


\bibitem[McCambridge et~al\mbox{.}(2014)]%
        {mccambridgeSystematicReviewHawthorne2014}
\bibfield{author}{\bibinfo{person}{Jim McCambridge}, \bibinfo{person}{John
  Witton}, {and} \bibinfo{person}{Diana~R. Elbourne}.}
  \bibinfo{year}{2014}\natexlab{}.
\newblock \showarticletitle{Systematic Review of the {{Hawthorne}} Effect: New
  Concepts Are Needed to Study Research Participation Effects}.
\newblock \bibinfo{journal}{\emph{Journal of Clinical Epidemiology}}
  \bibinfo{volume}{67}, \bibinfo{number}{3} (\bibinfo{date}{March}
  \bibinfo{year}{2014}), \bibinfo{pages}{267--277}.
\newblock
\showISSN{1878-5921}
\urldef\tempurl%
\url{https://doi.org/10.1016/j.jclinepi.2013.08.015}
\showDOI{\tempurl}


\bibitem[Randhavane et~al\mbox{.}(2019)]%
        {randhavaneIdentifyingEmotionsWalking2019}
\bibfield{author}{\bibinfo{person}{Tanmay Randhavane}, \bibinfo{person}{Uttaran
  Bhattacharya}, \bibinfo{person}{Kyra Kapsaskis}, \bibinfo{person}{Kurt Gray},
  \bibinfo{person}{Aniket Bera}, {and} \bibinfo{person}{Dinesh Manocha}.}
  \bibinfo{year}{2019}\natexlab{}.
\newblock \bibinfo{title}{Identifying {{Emotions}} from {{Walking}} Using
  {{Affective}} and {{Deep Features}}}.
\newblock
\newblock
\urldef\tempurl%
\url{https://doi.org/10.48550/arXiv.1906.11884}
\showDOI{\tempurl}


\bibitem[Rodriguez et~al\mbox{.}(2019)]%
        {rodriguezParameterizedAnimatedActivities2019}
\bibfield{author}{\bibinfo{person}{Alba M.~Rios Rodriguez},
  \bibinfo{person}{Steven Poulakos}, \bibinfo{person}{Maurizio Nitti},
  \bibinfo{person}{Mattia Ryffel}, {and} \bibinfo{person}{Robert Sumner}.}
  \bibinfo{year}{2019}\natexlab{}.
\newblock \showarticletitle{Parameterized {{Animated Activities}}}. In
  \bibinfo{booktitle}{\emph{Motion, {{Interaction}} and {{Games}}}}
  \emph{(\bibinfo{series}{{{MIG}} '19})}. \bibinfo{publisher}{{ACM}},
  \bibinfo{address}{{New York, NY, USA}}, \bibinfo{pages}{11:1--11:9}.
\newblock
\showISBNx{978-1-4503-6994-7}
\urldef\tempurl%
\url{https://doi.org/10.1145/3359566.3360062}
\showDOI{\tempurl}


\bibitem[Rose et~al\mbox{.}(1998)]%
        {roseVerbsAdverbsMultidimensional1998}
\bibfield{author}{\bibinfo{person}{Charles Rose}, \bibinfo{person}{Michael~F.
  Cohen}, {and} \bibinfo{person}{Bobby Bodenheimer}.}
  \bibinfo{year}{1998}\natexlab{}.
\newblock \showarticletitle{Verbs and Adverbs: Multidimensional Motion
  Interpolation}.
\newblock \bibinfo{journal}{\emph{IEEE Computer Graphics and Applications}}
  \bibinfo{volume}{18}, \bibinfo{number}{5} (\bibinfo{date}{Sept.}
  \bibinfo{year}{1998}), \bibinfo{pages}{32--40}.
\newblock
\showISSN{1558-1756}
\urldef\tempurl%
\url{https://doi.org/10.1109/38.708559}
\showDOI{\tempurl}


\bibitem[University(2003)]%
        {carnegiemellonuniversityCMUGraphicsLab2003}
\bibfield{author}{\bibinfo{person}{Carnegie~Mellon University}.}
  \bibinfo{year}{2003}\natexlab{}.
\newblock \bibinfo{title}{{{CMU Graphics Lab Motion Capture Database}}}.
\newblock
\newblock


\bibitem[Victor et~al\mbox{.}(2021)]%
        {victorLearningbasedPoseEdition2021}
\bibfield{author}{\bibinfo{person}{L{\'e}on Victor}, \bibinfo{person}{Alexandre
  Meyer}, {and} \bibinfo{person}{Sa{\"i}da Bouakaz}.}
  \bibinfo{year}{2021}\natexlab{}.
\newblock \showarticletitle{Learning-Based Pose Edition for Efficient and
  Interactive Design}.
\newblock \bibinfo{journal}{\emph{Computer Animation and Virtual Worlds}}
  \bibinfo{volume}{32}, \bibinfo{number}{3-4} (\bibinfo{year}{2021}),
  \bibinfo{pages}{e2013}.
\newblock
\showISSN{1546-427X}
\urldef\tempurl%
\url{https://doi.org/10.1002/cav.2013}
\showDOI{\tempurl}


\bibitem[von Laban and Ulmann(1971)]%
        {labanMasteryMovement1971}
\bibfield{author}{\bibinfo{person}{Rudolf von Laban} {and}
  \bibinfo{person}{Lisa Ulmann}.} \bibinfo{year}{1971}\natexlab{}.
\newblock \bibinfo{booktitle}{\emph{The {{Mastery}} of {{Movement}}}}.
\newblock


\bibitem[Wang and Chen(1991)]%
        {wangCCDInverseKinematics}
\bibfield{author}{\bibinfo{person}{Li-Chun~Tommy Wang} {and}
  \bibinfo{person}{Chih~Cheng Chen}.} \bibinfo{year}{Aug./1991}\natexlab{}.
\newblock \showarticletitle{A Combined Optimization Method for Solving the
  Inverse Kinematics Problems of Mechanical Manipulators}.
\newblock \bibinfo{journal}{\emph{IEEE Transactions on Robotics and
  Automation}} \bibinfo{volume}{7}, \bibinfo{number}{4}
  (\bibinfo{year}{Aug./1991}), \bibinfo{pages}{489--499}.
\newblock
\showISSN{1042296X}
\urldef\tempurl%
\url{https://doi.org/10.1109/70.86079}
\showDOI{\tempurl}


\bibitem[Wang et~al\mbox{.}(2020)]%
        {wangAdversarialLearningModeling2020}
\bibfield{author}{\bibinfo{person}{Qi Wang}, \bibinfo{person}{Thierry
  Arti{\`e}res}, \bibinfo{person}{Mickael Chen}, {and} \bibinfo{person}{Ludovic
  Denoyer}.} \bibinfo{year}{2020}\natexlab{}.
\newblock \showarticletitle{Adversarial Learning for Modeling Human Motion}.
\newblock \bibinfo{journal}{\emph{The Visual Computer}} \bibinfo{volume}{36},
  \bibinfo{number}{1} (\bibinfo{date}{Jan.} \bibinfo{year}{2020}),
  \bibinfo{pages}{141--160}.
\newblock
\showISSN{1432-2315}
\urldef\tempurl%
\url{https://doi.org/10.1007/s00371-018-1594-7}
\showDOI{\tempurl}


\bibitem[Wen et~al\mbox{.}(2021)]%
        {wenAutoregressiveStylizedMotion2021}
\bibfield{author}{\bibinfo{person}{Yu-Hui Wen}, \bibinfo{person}{Zhipeng Yang},
  \bibinfo{person}{Hongbo Fu}, \bibinfo{person}{Lin Gao},
  \bibinfo{person}{Yanan Sun}, {and} \bibinfo{person}{Yong-Jin Liu}.}
  \bibinfo{year}{2021}\natexlab{}.
\newblock \showarticletitle{Autoregressive {{Stylized Motion Synthesis With
  Generative Flow}}}. In \bibinfo{booktitle}{\emph{Proceedings of the
  {{IEEE}}/{{CVF Conference}} on {{Computer Vision}} and {{Pattern
  Recognition}}}}. \bibinfo{pages}{13612--13621}.
\newblock
\urldef\tempurl%
\url{https://doi.org/10.1109/CVPR46437.2021.01340}
\showDOI{\tempurl}


\bibitem[Wu et~al\mbox{.}(2011)]%
        {wuNaturalCharacterPosing2011}
\bibfield{author}{\bibinfo{person}{Xiaomao Wu}, \bibinfo{person}{Maxime
  Tournier}, {and} \bibinfo{person}{Lionel Reveret}.}
  \bibinfo{year}{2011}\natexlab{}.
\newblock \showarticletitle{Natural {{Character Posing}} from a {{Large Motion
  Database}}}.
\newblock \bibinfo{journal}{\emph{IEEE Computer Graphics and Applications}}
  \bibinfo{volume}{31}, \bibinfo{number}{3} (\bibinfo{date}{May}
  \bibinfo{year}{2011}), \bibinfo{pages}{69--77}.
\newblock
\urldef\tempurl%
\url{https://doi.org/10.1109/MCG.2009.111}
\showDOI{\tempurl}


\bibitem[Xia et~al\mbox{.}(2015)]%
        {xiaRealtimeStyleTransfer2015}
\bibfield{author}{\bibinfo{person}{Shihong Xia}, \bibinfo{person}{Congyi Wang},
  \bibinfo{person}{Jinxiang Chai}, {and} \bibinfo{person}{Jessica Hodgins}.}
  \bibinfo{year}{2015}\natexlab{}.
\newblock \showarticletitle{Realtime Style Transfer for Unlabeled Heterogeneous
  Human Motion}.
\newblock \bibinfo{journal}{\emph{ACM Transactions on Graphics}}
  \bibinfo{volume}{34}, \bibinfo{number}{4} (\bibinfo{date}{July}
  \bibinfo{year}{2015}), \bibinfo{pages}{119:1--119:10}.
\newblock
\showISSN{07300301}
\urldef\tempurl%
\url{https://doi.org/10.1145/2766999}
\showDOI{\tempurl}


\bibitem[Yoshizaki et~al\mbox{.}(2011)]%
        {yoshizakiActuatedPhysicalPuppet2011}
\bibfield{author}{\bibinfo{person}{Wataru Yoshizaki}, \bibinfo{person}{Yuta
  Sugiura}, \bibinfo{person}{Albert~C. Chiou}, \bibinfo{person}{Sunao
  Hashimoto}, \bibinfo{person}{Masahiko Inami}, \bibinfo{person}{Takeo
  Igarashi}, \bibinfo{person}{Yoshiaki Akazawa}, \bibinfo{person}{Katsuaki
  Kawachi}, \bibinfo{person}{Satoshi Kagami}, {and} \bibinfo{person}{Masaaki
  Mochimaru}.} \bibinfo{year}{2011}\natexlab{}.
\newblock \showarticletitle{An Actuated Physical Puppet as an Input Device for
  Controlling a Digital Manikin}. In \bibinfo{booktitle}{\emph{Proceedings of
  the {{SIGCHI Conference}} on {{Human Factors}} in {{Computing Systems}}}}
  \emph{(\bibinfo{series}{{{CHI}} '11})}. \bibinfo{publisher}{{Association for
  Computing Machinery}}, \bibinfo{address}{{New York, NY, USA}},
  \bibinfo{pages}{637--646}.
\newblock
\showISBNx{978-1-4503-0228-9}
\urldef\tempurl%
\url{https://doi.org/10.1145/1978942.1979034}
\showDOI{\tempurl}


\end{thebibliography}


\end{document}